\documentclass[twocolumn,showpacs,preprintnumbers,amsmath,amssymb]{revtex4}
\usepackage{amsmath,amsthm}
\usepackage{graphicx}
\usepackage{dcolumn}
\usepackage{bm}

\begin{document}
\title{Comment on "Observation of neutronless fusion reactions in picosecond laser plasmas"} 
\author{S. Kimura}
 \email{kimura@lns.infn.it}
\altaffiliation[Also at ]{Dipartimento di Fisica e Astronomia dell'Universita' di Catania, via Santa Sofia, 64, 95123 Catania, Italy}
\author{A. Anzalone}
\author{A. Bonasera}
 \affiliation{Laboratori Nazionali del Sud, INFN, via Santa Sofia, 62, 95123 Catania, Italy}

\date{\today}

\begin{abstract} 
The paper by Belyaev et al.~[Phys. Rev. E {\bf 72}, 026406 (2005)] 
reported the first experimental observation of alpha particles produced in the thermonuclear 
reaction $^{11}$B($p,\alpha$)$^{8}$Be induced by laser-irradiation on a $^{11}$B polyethylene 
(CH$_2$) composite target.
The laser used in the experiment is characterized by a picosecond pulse duration and a
peak of intensity of 2$\times$10$^{18}$ W/cm$^2$.
We suggest that both the background-reduction method adopted in their detection system and 
the choice of the detection energy region of the reaction products are possibly inadequate. 
Consequently the total yield reported underestimates the true yield. 
Based on their observation, we give an estimation of the total yield to be higher than their conclusion, i.e.,
of the order of 10$^5 \alpha$ per shot.
\end{abstract}
\maketitle

The observations of the thermonuclear reactions in a high-power laser pulse irradiated target 
is one of the hottest topics~\cite{ditmire,norreys,disdier,fritzler,izumi,habara,belyaev:026406}. 
The most investigated reaction is D($d,n$)$^3$He with a $Q$-value of 3.26 MeV. 
There have been studies using different characteristics of lasers irradiation on 
a wide variety of targets, solid CD$_2$ plastic~\cite{norreys,izumi,habara}, 
D$_2$-gas~\cite{fritzler} and deuterium-clusters~\cite{ditmire}.    
Since the reactions produce monochromatic neutrons, the spectroscopy of these neutrons 
gives important information on the ion acceleration mechanism in the laser-induced plasma. 

In the experiment recently carried out by a Russian group 
the yield of 10$^3$ $\alpha$-particles has been reported~\cite{belyaev:026406}, for the first time, 
in the laser-irradiation of a $^{11}$B+CH$_2$ composite target. 
Their experiment is important for a deep understanding of the ion acceleration mechanism 
in the laser-matter interaction.   
The experiment has been carried out by using a ``Neodymium'' laser facility with the pulse energy of up to 15~J, 
a laser wave length of 1.055~$\mu$m, and a pulse duration of 1.5~ps. Before the main pulse, there are three pre-pulses 
with relative intensities 10$^{-4}$, 10$^{-3}$ and 10$^{-8}$, with ps durations for the former two and with 
4~ns duration for the last one.

\begin{figure}
  \includegraphics[height=.3\textheight]{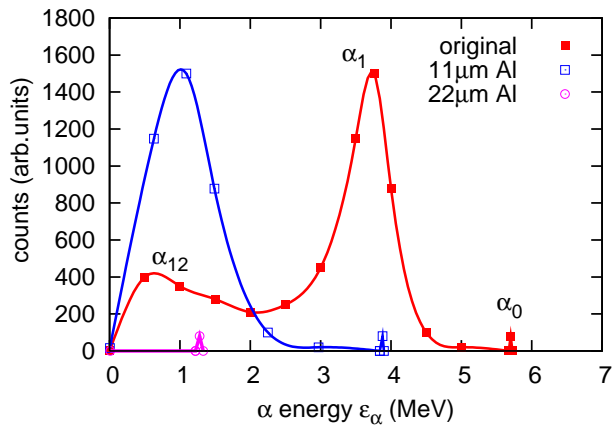}
  \caption{\label{fig:spec} Shifts of the $\alpha$ energy spectrum due to the 11~$\mu$m and 22~$\mu$m thick aluminum foils.
    The alpha energy spectrum in the reaction $^{11}$B($p,\alpha$)$^{8}$Be at 660~keV beam energy from Ref.~\cite{liu}  
  is given by full squares joined by the thick line. 
  The shifted spectra due to a 11~$\mu$m or 22~$\mu$m thick aluminum foils are given by the open squares and open 
  circles, respectively, joined by thick lines.   
  The small $\alpha_0$ sharp peak at 5.7~MeV in the original spectrum is broadened and shifted to 1.27~MeV 
  by the 22~$\mu$m thick Al foil. }
\end{figure}

The laser beam has been focused on the solid target at an oblique incidence of 40 degrees to the target normal.  
CR-39 track detectors covered with 11 and 22~$\mu$m thick aluminum foils have been used to count the yield of 
$\alpha$-particles from the reaction $^{11}$B($p,\alpha$)$^{8}$Be. 
The reaction induces three-particles decay. Either through the $^8$Be ground state~($\alpha_0$):  
\begin{equation}
  \label{eq:11bpa0}
\mathrm{^{11}B}+p \rightarrow \alpha_0+\mathrm{^8Be} 
\end{equation}
with the reaction $Q$-value~$= 8.59$~MeV 
or through the $^8$Be excited state~($\alpha_1$): 
\begin{equation}
  \label{eq:11bpa1}
\mathrm{^{11}B}+p \rightarrow \alpha_1+\mathrm{^8Be}^* 
\end{equation}
with the reaction $Q$-value~$= 5.65$~MeV and a large width of 1.5~MeV~\cite{ajzenberg,ajzenberg88,liu}. 
This is followed by the decay of the excited state~($\alpha_{12}$):
\begin{equation}
\mathrm{^8Be}^* \rightarrow 2 \alpha_{12}   \label{eq:11bpa12}
\end{equation}
and a reaction $Q$-value~$= 3.028$~MeV.
It is known that the main channel of the reaction is the second 
~\cite{be87,rauscher} and only 1 \%  
of the reaction products are $\alpha_0$ from the reaction ~(\ref{eq:11bpa0}). 
Using energy and momentum conservation laws, 
the $\alpha_0$ and $\alpha_1$ have kinetic energies: 
\begin{eqnarray}
  \varepsilon_{\alpha_0}&=&\frac{8}{12}(8.59+E) \ \mathrm{MeV} \\
  \varepsilon_{\alpha_1}&=&\frac{8}{12}(5.65+E) \ \mathrm{MeV} 
\end{eqnarray}
where $E$ is the center-of-mass incident energy in the case of the conventional beam-target experiment. 
But in the laser-induced plasma, the incident energy of the reactions is characterized by
some energy distributions, which are not known clearly. 
If we assume a thermal equilibrium state for the plasma, the energy distribution is given by a Maxwellian.
The temperature of the plasma is estimated~\cite{forslund,forslund2} to be of the order of 
67~keV for a background electron temperature $T_c$= 0.5~keV and 84~keV for $T_c$= 1.~keV
at the given laser intensity and the wavelength of the experiment.  
We mention that Ref.~\cite{krainov} gives an estimate of the nuclear temperature of 33~keV, lower than our estimation.  
The ions, therefore, can be accelerated up to the energies of the order of hundreds of keV at most.
At such low energies, the
$\alpha_0$ and $\alpha_1$ are estimated to have energies 5.7~MeV and 3.76~MeV, respectively, in the exit channel.
However, the energy spectrum of $\alpha_1$ has a large width, $\Gamma$= 1.5~MeV, consequently 
the $\alpha_{12}$ spectra spread from 0 to higher than 5~MeV~\cite{beckman}.  
An $\alpha$ energy spectrum obtained experimentally in Ref.~\cite{liu} shows clearly these characteristics 
of the reaction $^{11}$B($p,\alpha$)$^{8}$Be.  
The full squares connected by the thick line in Fig.~\ref{fig:spec} reproduce the data reported in 
Ref.~\cite{liu}. The two peaks at 3.76~MeV and 5.7~MeV are clearly visible.  

In the experiment in Ref.~\cite{belyaev:026406} 
the CR-39 track detectors have been placed at angles of 0, 45 and 85 degrees to the target normal. 
The $^{11}$B($p,\alpha$)$^{8}$Be reaction yield has been estimated by subtracting the background
obtained in the irradiation of the pure CH$_2$ target.  
The detectors are covered with Aluminum foils 11 or 22~$\mu$m thick.  
The reason for covering the plastic detectors is that the alpha tracks get confused with energetic ions 
coming from the high momentum tail of the plasma distributions. Cutting off the track diameter below 7~$\mu$m
as in Ref.~\cite{belyaev:026406} eliminates all the protons but not heavier ions ($\mathrm{B}$ and $\mathrm{C}$) of the plasma 
which leave bigger tracks. The authors of Ref.~\cite{belyaev:026406} observed that these 'strange ions' 
were still dominant when a 6~$\mu$m Al foil is used. This fact prompted them to increase the thickness
of the foil which caused blocking lower energy $\alpha$-particles as well.  
However this shielding of background is efficient, only if the energy of the detected ions is well specified 
as in the case of the reaction with a two-body exit-channel. By contrast in the reaction 
$^{11}$B($p,\alpha$)$^8$Be, the energy spectrum of reaction products spreads from 0 up to 5.7 MeV, as it is shown 
in Fig.~\ref{fig:spec}. 
In such a case the Al foil will remove the major part of the reaction products. 
A 11~$\mu$m thick Al foil shields $\alpha$-particles with energies lower than 3~MeV. 
If one uses a 22~$\mu$m thick Al foil, $\alpha$-particles with energies lower than 5~MeV will be stopped 
inside the foil.
We have performed simulations of the transmitted $\alpha$-particles through the foils by using TRIM in SRIM codes~\cite{srim}. 
Fig.~\ref{fig:spec} shows the $\alpha$-particle reduction by Al foils with thicknesses 
11~$\mu$m~(open squares) and 22~$\mu$m~(open circles) together with the original spectrum~(full squares). 
One can see clearly that the peak at 3.76~MeV is shifted and broadened passing through the 11~$\mu$m thick Al foil:
450 counts of $\alpha$-particles at the initial energy 3~MeV are reduced to 16 counts at 4~keV.
Using a 22~$\mu$m thick Al foil gives exclusively the $\alpha_0$ peak.       
Thus the fusion yield in Ref.~\cite{belyaev:026406} underestimates the true yield.  
Considering the expected energies of the reaction products from all the reaction channels, 
there is ample room for further improvement of the choice of this detection energy region.  

\begin{figure}
  \includegraphics[height=.3\textheight]{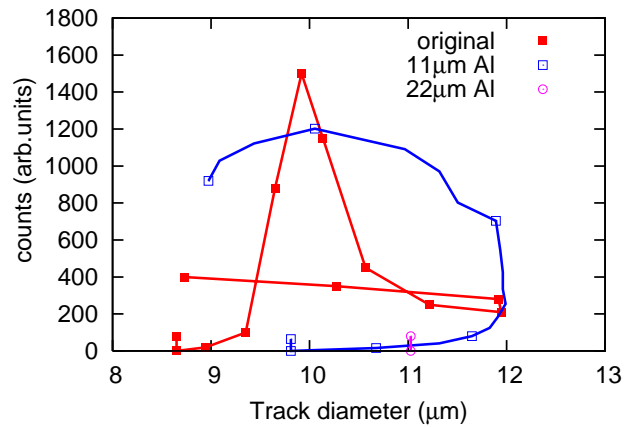}
  \caption{\label{fig:diam} Expected track diameters on CR-39 detectors from the original and shifted 
    $\alpha$-spectra. The symbols are as in Fig.~\ref{fig:spec}.}
\end{figure}
From the calibration data of the detectors by $\alpha$ sources, which is shown in Fig.~1 of Ref.~\cite{belyaev:026406},  
it is possible to convert the $\alpha$-energy spectrum in Fig.~\ref{fig:spec}
to the one as a function of observed track diameters.  
In Fig.~\ref{fig:diam} we show the $\alpha$-energy spectrum of Fig.~\ref{fig:spec} 
as a function of the track diameters on the CR-39 detectors.    

The 22~$\mu$m thick Al foil shields the major part of the reaction products and 
gives the 5.7~MeV $\alpha_0$ only which is a less important channel.   
In Fig.~\ref{fig:spec} one sees that the 
$\alpha_0$ will lose its energy passing through the foil 
and the transmitted $\alpha_0$ will have an energy of about 1.27~MeV$\pm$0.08~MeV.
Fig.~\ref{fig:diam} shows that  
the $\alpha$-particles in this energy range will give track diameters of about 11~$\mu$m.
Looking at Fig.~4 in Ref.~\cite{belyaev:026406}, which shows the distributions of track diameters
for detectors covered with 22~$\mu$m Al, an excess of tracks above the background with diameters around 
11~$\mu$m is, indeed, recognized.   
If this estimation is correct, we can conclude that the yield of $\alpha_0$ is 
about 1.5$\times$10$^3$/4$\pi$~Sr, from the values tabulated in their table I, 
under the assumption of an isotropic distribution of the reaction products.    

\begin{figure}
  \includegraphics[height=0.45\textheight]{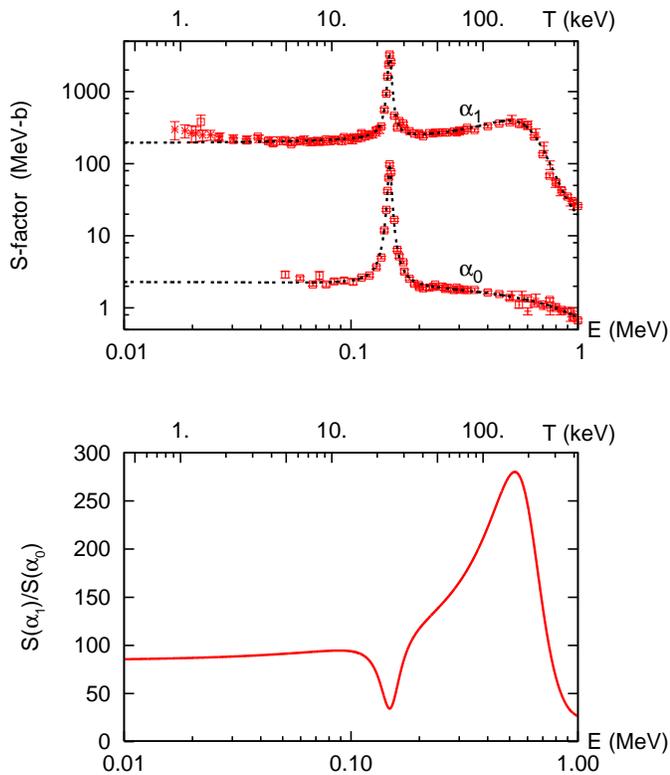}
  \caption{\label{fig:sfact} The $S$-factors for the reaction $^{11}$B($p,\alpha$)$^{8}$Be as a function of the incident center-of-mass energy~(the abscissa) and of the plasma temperature~(the mirror of the abscissa) for two reaction channels~(top panel). Experimental data are taken from~\cite{be87}~(squares) and~\cite{an93}~(crosses). The dotted fitting curves to the data are obtained using polynomial expression for the non-resonant contribution and Breit-Wigner formula for the resonant contribution. In the low energy region the screening effect due to bound electrons is not included. In the bottom panel, the ratio of the S-factors in the $\alpha_1$ and $\alpha_0$ channels is shown.}
\end{figure}

From this result, 
we might be able to estimate the true fusion yield
by considering the ratio of the astrophysical $S$-factors~\cite{clayton}, which is directly related to the reaction cross section, 
in the reaction~(\ref{eq:11bpa1}) to the reaction~(\ref{eq:11bpa0})
~\cite{rauscher}. 
Fig.~\ref{fig:sfact} shows the $S$-factors for the two channels and their ratio
both as functions of the incident center-of-mass energy $E$ of colliding nuclei and the plasma 
temperature $T$. 
The correspondence between $E$ and $T$ is obtained by the relation between the plasma temperature 
and the so-called ``most effective energy'' at that temperature~\cite{clayton, nacre}.  
For the purpose of taking the ratio, the experimental data of the $S$-factors~\cite{be87,an93} 
have been fitted by a polynomial expression combined with the Breit-Wigner resonance formula. 
The resulting curves, as well as the experimental data, are shown by the dashed lines in the top panel. 
In the bottom panel of Fig.~\ref{fig:sfact}, the curve shows the ratio of the $S$-factors.
The ratio is almost constant up to the temperature 20~keV but varies from 30 to 280 
in the temperature region where two resonances at $E=$148.5~keV and 660~keV dominate the $S$-factors.    
The increase of the ratio is attributed to the presence of a broad resonance at 660~keV exclusively in the 
$\alpha_1$ channel. 
Provided that the ratio of the $S$-factors
is from 100 to 170 in the range of the plasma temperature from 33 to 84~keV, 
the total fusion yield is estimated to be about 130 
times of the observed value with 22~$\mu$m Al foil, i.e., more than 2.$\times$10$^5$~fusions per shot.
This value becomes 4.$\times$10$^4$, if we use the averaged value over bursts reported in Ref.~\cite{belyaev:026406}.

\bigskip
In conclusion we have
discussed the expected energy range of the reaction products from 
the thermonuclear reaction 
$^{11}$B($p,\alpha$)$^{8}$Be induced by an irradiation on a $^{11}$B polyethylene composite target, 
whose first quantitative observation has been given 
by Belyaev et al.~\cite{belyaev:026406}. 
Their experiment is essential not only to seek a possibility of aneutronic fusions
but also to promote a better understanding of the ion acceleration mechanism 
in the laser-matter interaction. In this connection, it is highly desirable that more 
precise measurements of the angular distribution of the reaction products
will be performed~\cite{abc}. 
We have demonstrated that  
the observed yield in their experiment is underestimated at least by a factor of 100, 
due to both the background reduction method in their detection system and 
their selection of the detection energy region.                 
   
\bigskip
This work was partially supported by the research project "Structure and Dynamics of nuclear 
systems beyond the limits of stability" directed by Prof. U. Lombardo and financed by Department
of physics and astronomy of the University of Catania.  
One of us~(S. K.) would like to express her gratitude to Prof. U. Lombardo for the financial support 
from the University of Catania. 
She acknowledges Dr. M.R.D. Rodrigues for fruitful discussions.

\bibliography{memo.bib}
%



\end{document}